\documentclass[pra,twocolumn,superscriptaddress,showpacs,preprintnumbers,amsmath,amssymb]{revtex4-2}
\usepackage{graphicx}
\usepackage{dcolumn}
\usepackage{bm}
\usepackage{enumerate}
\usepackage{mathrsfs}
\usepackage{amsmath}
\usepackage{changes}
\usepackage{lipsum}
\usepackage{booktabs, array, float, tabularx, booktabs, lipsum}
\usepackage{makecell}
\definechangesauthor[name={Author1}, color=red]{A1}
\definechangesauthor[name={Author2}, color=green]{A2}
\usepackage{epstopdf}
\usepackage[titletoc]{appendix}
\usepackage{makecell}
\usepackage{braket}
\begin{document}
	
\title{Heisenberg-limit spin squeezing with spin Bogoliubov Hamiltonian}
	
	\author{Jun Zhang}
	\affiliation{Key Laboratory of Artificial Micro- and Nano-structures of Ministry of Education,
		and School of Physics and Technology, Wuhan University, Wuhan, Hubei 430072, China}
	
	\author{Sheng Chang}
	\affiliation{Key Laboratory of Artificial Micro- and Nano-structures of Ministry of Education,
		and School of Physics and Technology, Wuhan University, Wuhan, Hubei 430072, China}
	
	
	\author{Wenxian Zhang}
	\email[Corresponding email:]{wxzhang@whu.edu.cn}
	\affiliation{Key Laboratory of Artificial Micro- and Nano-structures of Ministry of Education,
		and School of Physics and Technology, Wuhan University, Wuhan, Hubei 430072, China}
	\affiliation{Wuhan Institute of Quantum Technology, Wuhan, Hubei 430206, China}
	
	\date{\today}
\begin{abstract}
It is well established that the optimal spin squeezing under a one-axis-twisting Hamiltonian follows a scaling law of $J^{-2/3}$ for $J$ interacting atoms after a quench dynamics. Here we prove analytically and numerically that the spin squeezing of the ground state of the one-axis-twisting Hamiltonian actually reaches the Heisenberg limit $J^{-1}$. By constructing a bilinear Bogoliubov Hamiltonian with the raising and lowering spin operators, we exactly diagonalize the spin Bogoliubov Hamiltonian, which includes the one-axis twisting Hamiltonian as a limiting case. The ground state of the spin Bogoliubov Hamiltonian exhibits wonderful spin squeezing, which approaches to the Heisenberg limit in the case of the one-axis twisting Hamiltonian. It is possible to realize experimentally the spin squeezed ground state of the one-axis-twisting Hamiltonian in dipolar spinor condensates, ultracold atoms in optical lattices, spins in a cavity, or alkali atoms in a vapor cell.
\end{abstract}
	\maketitle
\section{Introduction}
 Development of the methods for realizing squeezed spin states (SSS) or entangled spin states has been a vigorous frontier in quantum enhanced precision measurement for more than three decades~\cite{Wineland1992,Kitagawa1993,Wineland1994}, because such strongly correlated quantum states may lead to significant improvement in atomic clock~\cite{Leroux2010atomclock,Ludlow2015RMP,Hosten2016Nature,Pedrozo2020Nature,Schulte2020NC}, optical or atomic interferometer~\cite{Pedrozo2020Nature,Szigeti2020PRL,Genovese2021AVS}, frequency standards~\cite{Huelga1997PRL,Shaniv2018PRL}, and gravitational wave detection~\cite{McKenzie2002PRL,Aasi2013NPhot,Szigeti2020PRL}. The uncertainty principle predicts that relative measurement precision approaches to standard quantum limit (SQL), $J^{-1/2}$ for uncorrelated $J$ particles in a quantum interferometer, and Heisenberg limit (HL), $J^{-1}$ for some special squeezed or entangled quantum states~\cite{RMPSmerzi}. Generating atomic spin squeezing has been proposed and demonstrated with a variety of methods, including quantum nondemolition measurements~\cite{Kuzmich_1998}, squeezing transferring~\cite{Hald1999PRL,Lukin2000PRL,Huang2021NPJ}, and dynamical evolution~\cite{Wineland1992, Wineland1994, Kitagawa1993}, in systems such as atom-cavity interaction systems~\cite{Ueda2003PRA, Takahashi2005PRL, Leroux2010PRL}, interacting trapped ions~\cite{Wineland1996PRL, Justin2016Science}, and Bose-Einstein condensates (BECs)~\cite{Esteve2008,Maussang2010,Riedel2010, Hamley2012,Berrada2013,Strobel2014Science, Dicke2014PRL}.

Nonlinear atomic interaction induces spin squeezing after a quench dynamics from a coherent spin state (CSS). It is generally believed that the spin squeezing parameter of $J$ interacting atoms may achieve $J^{-2/3}$ under a one-axis-twisting (OAT) Hamiltonian and $J^{-1}$ under a two-axis-twisting (TAT) Hamiltonian~\cite{Kitagawa1993}. However, the SSSs generated with the dynamical evolution slip away quickly after the optimal time. The long-lived ground SSSs are thus desired and have been explored recently in a spinor BEC by Chapman's group~\cite{Chapman2023} and in a trapped-atom clock system by Reichel's group~\cite{Huang2023PRXQuantum}.

We investigate in this Letter the ground SSS of a spin Bogoliubov Hamiltonian. By constructing a bilinear Bogoliubov Hamiltonian with spin operators, which includes the OAT Hamiltonian as a limiting case, we prove analytically and numerically that the spin squeezing of the ground SSS of the spin Bogoliubov Hamiltonian (as well as the OAT Hamiltonian) reaches the HL $J^{-1}$ with a weak external field. Our theoretical results may be realized potentially in many experiments, such as dipolar spinor BEC~\cite{Law98,Zoller2001nature,Zoller2002PRA,You2002PRA,Zoller2003PRA,Yi2004PRL,Meystre2004PRA,Pezz2005PRA,Choi2005PRA,Yi2006PRA,You2007PRA,Yi2012PRA}, ultracold atoms in an optical lattice\cite{Sorensen1999PRL,Gerbier2006PRL,MA2011,Pedrozo2020Nature,Plodzien2020PRA,Yanes2022PRL,Bornet2023Nature,Dziurawiec2023PRA}, and spins in a cavity or in a vapor cell~\cite{Agarwal1997PRA,Dimer2007PRA,Morrison2008PRA, Morrison2008PRL, Leroux2010PRL, Bennett2013PRL,Dalla2013PRL,Masson2017PRL, Zhiqiang2017OPtica, Borregaard2017NJP,Groszkowski2020PRL, Groiseau2021PRA,Huang2021NPJ,Groszkowski2022PRL,Li2022PRXQuantum,Huang2023PRA}.

\section{LMG model}

 The Lipkin-Meshkov-Glick (LMG) model originally introduced in nuclear physics in 1965, has been widely studied in many quantum spin systems, such as spinor BEC, Rydberg-dressed atomic gas and cold atoms in optical lattices~\cite{Gil2014PRL,Zeiher2016NP,Muniz2020Nature, Borish2020PRL,Defenu2023RMP}. The entanglement properties as well as spin squeezing of this model in its ground state have been discussed in many literatures with mean field theory~\cite{Vidal2004PRL, VidalPRB2005, Ma2009PRA}, by treating the quantum effect as small fluctuations and using the Holstein-Primakoff transformation in the thermodynamic limit~\cite{HP1940PR}. Different from previous methods, we explore an exact analytical approach to diagonalize the Hamiltonian for a certain parameters and focus on the spin-squeezed ground state.

The Hamiltonian of the LMG model reads
\begin{eqnarray}\label{Eq:LMGmodel}
	H=\eta(J_x^2+\gamma J_y^2)+\Omega J_z,
\end{eqnarray}
where $J_{x,y,z}$ are collective spin operator, $\eta$ the interaction strength, $\gamma$ the anisotropic parameter, and $\Omega$ the effective external field. At zero external field $\Omega = 0$, this Hamiltonian becomes an OAT Hamiltonian $H_{OAT}\propto J_{x,z}^2$ if $\gamma=0, 1$ and a TAT Hamiltonian $H_{TAT}\propto (J_x^2-J_{y,z}^2)$ if $\gamma = -1, 1/2$, after dropping the constant term with ${\mathbf J}^2 = J_x^2 + J_y^2 + J_z^2$.

\section{Construction of Bogoliubov Hamiltonian}
 To diagonalize the Hamiltonian Eq.~(\ref{Eq:LMGmodel}), we construct a diagonal ``Bogoliubov" Hamiltonian $H_B$ through the Bogoliubov transformation of the spin lowering/raising operator $J_{\pm}=J_x \pm iJ_y$, inspired by the concept of squeezed light~\cite{gerry_knight_2004}. We prove that the constructed Bogoliubov Hamiltonian $H_B$ is a special case of the LMG model Eq.~(\ref{Eq:LMGmodel}).

Spin lowering and raising operators obey the canonical commutation relation,
$[J_+,\;J_-]=2J_z$ and $[J_{\pm},\; J_z]= \mp J_{\pm}$. A Dicke state $\ket{J,M}$ ($M=-J,-J+1,\cdots, J$) is the eigenstate of the spin operator $J_z$, satisfying the relation $J_z\ket{J,M}=M\ket{J,M}$. Since $J_z=\frac{1}{2}(J_+J_{-}-J_{-}J_+)$, a Dicke state is
\begin{eqnarray} \ket{J,M}&=&\frac{1}{(J+M)!}\left(\frac{2J}{J+M}\right)^{1/2}J_{+}^{J+M}\ket{J,-J}. \nonumber
\end{eqnarray}
Unlike the annihilation operator $\hat{a}$ of a photon, the spin lowering operator $J_-$ has only one eigenstate $\ket{J,-J}$, which is the vacuum state and satisfies $J_-\ket{J,-J}=0$.

Similar to generalized creation and annihilation operators, we define the following generalized spin lowering and raising operators,
$A_+=\mu^* J_{+}-\nu J_{-}, A_-=A_{+}^{\dagger}=\mu J_{-}-\nu^{\ast}J_+$.
To maintain the commutation relation $[J_+,\; J_{-}]=[A_+,\;A_{-}]$, we require
$
	\mid \mu\mid^2-\mid \nu\mid^2=1
$. We construct the following diagonal Hamiltonian using $A_{\pm}$, the Bogoliubov Hamiltonian,
\begin{eqnarray}
	H_B &=& A_{+}A_{-}.
\end{eqnarray}
By assuming $\mu=\cosh\theta$ and $\nu=e^{i\varphi}\sinh\theta$, one finds
$
	H_B
	=J_z + \cosh{2\theta}(J_x^2+J_y^2)
	-\cos{\varphi}\sinh{2\theta}(J_x^2-J_y^2)
	-\frac{i}{2}\sin{\varphi}\sinh{2\theta}(J^2_{+}-J^2_{-})
$. When $\varphi=\pi$, we obtain a specific spin Bogoliubov Hamiltonian as
\begin{eqnarray}\label{Eq:Ham_sq1}
	H_B&=&J_z + \cosh{2\theta}(J_x^2+J_y^2) + \sinh{2\theta}(J_x^2-J_y^2).
\end{eqnarray}
It is obvious that $H_B$ in Eq.~(\ref{Eq:Ham_sq1}) is a special case of the LMG Hamiltonian Eq.~(\ref{Eq:LMGmodel}), where $\gamma = \exp(-4\theta)$ and $\eta/\Omega = \exp(2\theta)$ must be satisfied. For a large enough $\theta$, the Bogoliubov Hamiltonian becomes effectively the OAT Hamiltonian, $H_{OAT} = \exp(2\theta) J_x^2$.

\subsection{Spin vacuum state}

The ground state of the Bogoliubov Hamiltonian $H_B$ is the eigenstate of the operator $A_{-}$,which is a spin vacuum state satisfying
\begin{eqnarray}\label{Eq:gschi}
	A_{-}\ket{\chi}&=&0\ket{\chi}.
\end{eqnarray}
This state can be expanded in the Dicke basis as $\ket{\chi}=\sum\limits_{M=-J}^{J}C_M \ket{J,M}$. By substituting the definition of $A_-$, we find the following recursion relation
\begin{eqnarray}
	C_{M+2}=\frac{\nu^{\ast}}{\mu}C_{M} \sqrt{\frac{(J-M)(J+M+1)}{(J-M-1)(J+M+2)}}.
\end{eqnarray}
It is straightforward to calculate the general formula in terms of the coefficient $C_{-J}$,
\begin{eqnarray}~\label{Eq.recur}
	C_{-J+2K}&=&\left(\frac{\nu^{\ast}}{\mu}\right)^K\frac{J!}{(J-K)!K!}
	\sqrt{\frac{(2J-2K)!(2K)!}{(2J)!}}C_{-J}, \nonumber\\	
\end{eqnarray}
$K=0,1,\cdots,J$. Since $\bra{J,J}A_{-}\ket{\chi}=0$ and $C_{J-1}=0$, it is easy to find that $C_{-J+2K+1}=0$ for $K=0,1,\cdots,J-1$. By further utilizing the normalization condition $\sum_{K=0}^{J}|C_{-J+2K}|^2=1$, we obtain
\begin{eqnarray}~\label{Eq.coff}
	C_{-J}&=&\frac 1 {\sqrt{\mathcal{F}\left(
	\frac{1}{2},-J,\frac{1}{2}-J, \left|\frac{\nu}{\mu}\right|^2\right)}}
\end{eqnarray}
where $\mathcal{F}[a,b,c,x]$ represents the hypergeometric function.

\subsection{Spin squeezing of the spin vacuum state}

Similar to a photon squeezed state, the spin vacuum state is also spin squeezed. To characterize the spin squeezing, two kinds of spin squeezing parameters were introduced by Kitagawa \textit{et al.} and Wineland \textit{et al.} respectively~\cite{Kitagawa1993, Wineland1994},
\begin{eqnarray}
	\xi_S^2=\frac{(\Delta J_{n_\perp})^2}{(J/2)}, \quad \xi_R^2=\frac{2J(\Delta J_{n_\perp})^2}{|\langle \vec{J}\rangle |^2}
	\label{Eq.para}
\end{eqnarray}
where subscript $n_{\perp}$ refers to an arbitrary axis perpendicular to the mean spin $\langle\vec{J}\rangle$, where the minimum of $(\Delta J_{n_\perp})^2$ is obtained. The inequality $\xi_{i}^2<1\;(i=S,R)$ indicates that the state is spin squeezed, compared to a CSS with $\xi_i^2 = 1$. The relation between the two squeezing parameters is $\xi_R^2=({J}/{|\langle \vec{J}\rangle|})^2\xi_S^2$. Since $J\geq|\langle \vec{J}\rangle |$ always holds, one finds $\xi_S^2 \leq \xi_R^2$~\cite{MA2011}. The equality is taken at $J=|\langle \vec{J}\rangle |$, when a CSS is considered. The spin squeezing parameter $\xi_R^2$ relates directly to the phase sensitivity of a quantum interferometer, which takes the general form as
\begin{eqnarray}
	\Delta\phi&=&\frac{(\Delta J_{n_\perp})}{|\langle \vec{J}\rangle |}
	=\frac{\xi_R}{\sqrt{J}}.
	\label{Eq.sensi}
\end{eqnarray}
Obviously, the phase sensitivity of a CSS is, $\Delta\phi={1}/{\sqrt{J}}$, which is denoted as the SQL~\cite{RMPSmerzi}. By contrast, an SSS is expected to achieve a phase sensitivity below the SQL but above the HL, i.e., $1/J \le \Delta\phi < 1/\sqrt{J}$~\cite{RMPSmerzi}. It follows immediately that $1/J \le \xi_R^2 <1$. On the other hand, $\xi_S^2$ may approach to zero, not constrained by the HL. For instance, a Dicke state $\ket{J,M \neq 0}$ has a constant spin size $J$ and zero spin variance along $z$-direction, resulting in $\xi_S^2=0$.

To calculate the parameters $\xi_S^2$ and $\xi_R^2$ of the spin vacuum state, we need to find spin average $\langle J_{n}\rangle$ and the minimal variance perpendicular to the mean-spin direction (MSD), i.e. $\Delta J_{n_\perp}$. Since the operators $J_{x,y}$ are a linear combination of $A_{\pm}$, it follows immediately that $\bra{\chi}J_{x,y}\ket{\chi} = 0$. By further employing the commutator $[A_+,A_-]=2J_z$, we find that $\bra{\chi}J_z\ket{\chi} \neq 0$ thus the MSD of the state $\ket{\chi}$ is along $z$-direction. The minimal variance must be in the $x$-$y$ plane and the covariance matrix is defined as
\begin{equation}
	\Gamma_{xy}=\left(\begin{array}{cc}
		\langle J^2_x\rangle & Cov(J_x,J_y)\\
		Cov(J_x,J_y)&\langle J^2_y\rangle \\
	\end{array}
	\right),
\end{equation}
with $Cov(J_x,J_y)=\left(\langle [J_x, J_y]_{+}\rangle-\langle J_x\rangle \langle J_y\rangle\right)/2$ being the covariance between $J_x$ and $J_y$, and
$[X, Y]_{+}=XY+YX$ denoting the anti-commutator. The eigenvalues of the covariance matrix are
\begin{eqnarray}
	\lambda_{\pm}&=&\frac{1}{2}\left[\langle J_x^2+J_y^2\rangle\pm\sqrt{\langle J_x^2-J_y^2\rangle^2+4Cov(J_x,J_y)^2}\;\right].\nonumber\\
\end{eqnarray}
After a straightforward calculation the eigenvalues are simplified as
$
\lambda_{\pm}=\frac{1}{2}\left[(F-K)\pm\sqrt{(F-K)^2-G^2}\right]
$
where $F=J(J+1)$, $G=\langle J_z\rangle$ and $K=\langle J^2_z\rangle$. Obviously we find that $min (\Delta J_{n_\perp})^2 =\lambda_{-}$~\cite{MA2011}. The squeezing parameters become
\begin{eqnarray}\label{Eq:xi_rs}
	\xi_S^2&=&\frac{2\lambda_{-}}{J}, \nonumber\\
	\xi_R^2&=&\frac{2J\lambda_{-}}{|\langle J_z\rangle |^2} = \frac{J}{2\lambda_+}.
\end{eqnarray}

In the limit $\theta\rightarrow\infty$, we approximate $(\nu^*/\mu)^K \approx 1-2K\exp(-2\theta)$ and the spin average becomes~\cite{approx}
\begin{eqnarray}
\label{Eq:Jzavg}
\langle J_z\rangle \approx -J^2 \exp(-2\theta).
\end{eqnarray}
Accordingly, the eigenvalues of covariance matrix are
$
\lambda_+ \approx   J^2/2
$ and
$
\lambda_- \approx (J^2/2) \exp(-4\theta).
$
Finally, we find that the squeezing parameters are
\begin{eqnarray}\label{Eq:limval}
 \xi_R^2&\approx &\frac{1}{J}, \quad
 \xi_S^2\approx J \exp(-4\theta).
\end{eqnarray}
In addition, the Bogoliubov Hamiltonian in Eq.~(\ref{Eq:Ham_sq1}) reduces effectively to the OAT Hamiltonian in the limit of $\theta \rightarrow \infty$. Therefore, we prove that the spin squeezing of the ground state for an OAT Hamiltonian is approaching HL, surpassing the constraint of $\propto J^{-2/3}$ for dynamic evolution governed by the same Hamiltonian. Moreover, the resulting ground state of $H_B$ is categorized as generalized intelligent state (GIS) which minimize the Robertson-Schr\"{o}dinger uncertainty relation~\cite{Robertson1930general,schrodinger1930,Kinani_2001GIS}. It is  straightforward to find that $\lambda_{+}\lambda_{-}=G^2/4=\langle J_z\rangle^2/4$, indicating that $\ket{\chi}$ is an intelligent state~\cite{note1}.

\begin{widetext}
		
\begin{figure}[th]
	\centering
	\includegraphics[width=0.8\columnwidth]{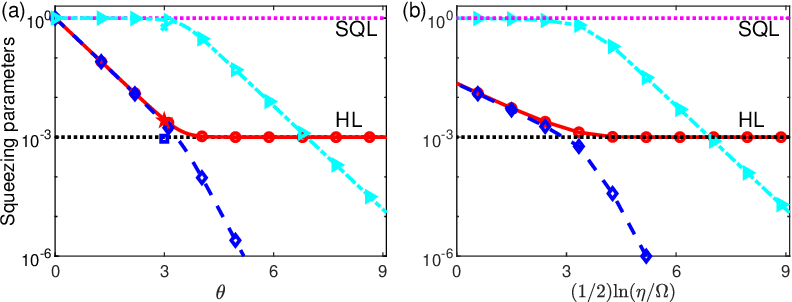}
	\caption{Squeezing parameters $\xi_S^2$ and $\xi_R^2$ of the ground state of (a) the spin Bogoliubov Hamiltonian $H_{B}$ Eq.~(\ref{Eq:Ham_sq1}) and (b) the anisotropic LMG Hamiltonian $H_A$. In (a), the spin squeezing parameters, $\xi_R^2$, $\xi_S^2$ and spin average $|\langle J_z\rangle|/J$  of state $\ket{\chi}$ are denoted by red, blue and cyan solid lines respectively. The same values of the ground state that calculated numerically by directly diagonalizing the Hamiltonian $H_B$ are denoted by red circles, blue diamonds and cyan triangles. The numerical and analytical results agree exactly with each other, indicating that the spin vacuum state $\ket{\chi}$  is indeed the ground state of the Hamiltonian $H_B$.
For large values of $\theta$, $\xi_R^2$ approaches to HL(lower black dotted line), while $\xi_S^2$  decreases exponentially to zero. The spin average $|\langle J_z\rangle|/J$ also decreases exponentially to zero. For the SQL (upper pink dotted line) $\xi_{S,R}^2 = 1$ and for the HL $\xi_R^2 = 1/J$. The cyan cross, pink pentagram and blue square denote respectively the spin average, $\xi_R^2$, and $\xi_S^2$ for the optimal SSS generated through dynamical evolution under the TAT Hamiltonian Eq.~(\ref{Eq:LMGmodel}) with $\gamma = -1, \Omega = 0$. Similarly, in (b) $\xi_R^2$ (red solid line with circles) for the ground state of the OAT Hamiltonian (at a large value of $\eta/\Omega$) approaches to the HL. Accordingly, $\xi_S^2$ (blue dashed line with diamonds) and $|\langle J_z\rangle|/J$ (cyan dash-dotted with triangles) decrease exponentially. The spin size is $J=1000$.}
	\label{fig.xi_rs}
\end{figure}

\end{widetext}

We present in Fig.~\ref{fig.xi_rs}(a) the squeezing parameters $\xi_R^2$ and $\xi_S^2$, as well as the spin average $\langle J_z\rangle/J$, of the state $\ket{\chi}$ for a spin size $J=1000$. These quantities vary with the parameter $\theta$. The coefficients $C_M$ of the ground state $\ket{\chi}$ are analytically calculated using the recursion relation Eq.~(\ref{Eq.recur}) and the coefficient $C_{-J}$. It is then straightforward to calculate the spin average $\langle \chi| J_z|\chi\rangle$ and the spin squeezing parameters $\xi_R^2$ and $\xi_S^2$. We also present the spin squeezing parameters $\xi_R^2$ and $\xi_S^2$ of the optimal SSS, which is generated numerically by evolving the system from an initial CSS $|J,J\rangle$ under the TAT Hamiltonian $H_{TAT} = J_x^2-J_y^2$~\cite{Kitagawa1993}. As a comparison, we plot in Fig.~\ref{fig.xi_rs}(b) the same quantities of the ground state of a special anisotropic LMG Hamiltonian with $\gamma =0$, $H_{A} = \eta J_x^2 + \Omega J_z$. As $\eta\gg \Omega$, the Hamiltonian $H_A$ effectively reduces to the OAT Hamiltonian. Because it is challenging with analytical method, the ground state of $H_{A}$ is calculated numerically for $J=1000$.

As shown in Fig.~\ref{fig.xi_rs}(a), the spin squeezing parameter $\xi_R^2$ approaches to the HL quickly in the large $\theta$ region ($\theta > 3$). The parameter $\xi_S^2$ and the spin average $\langle J_z\rangle$ decrease exponentially to zero approximately described by Eqs.~(\ref{Eq:Jzavg}) and (\ref{Eq:limval}), respectively. In the small $\theta$ region ($\theta < 3$), the spin average is almost a constant while the spin squeezing parameters $\xi_{S,R}^2$ decrease exponentially. This feature is especially useful for quantum enhanced metrology~\cite{Giovannetti2004Science,Giovannetti2006PRL,Giovannetti2011NP,MA2011,Toth2014,RMPSmerzi}. At $\theta \approx 3$, the squeezing parameters $\xi_{S,R}^2$ of the ground state $|\chi\rangle$ and of the optimal SSS of $H_{TAT}$ are close to the HL. On the other hand, the spin averages are quite different, $\langle \chi| J_z |\chi\rangle = -0.89J$ and $ |\langle SSS| J_z |SSS\rangle = 0.59J$. Similarly in Fig.~\ref{fig.xi_rs}(b), the spin squeezing parameters $\xi_R^2$ also approaches to the HL as $\eta/\Omega$ becomes large. Both panels (a) and (b) indicate that the spin squeezing of the ground state of the OAT Hamiltonian is at the HL, in stark contrast to earlier understanding~\cite{MA2011, Zhang_2014NJP}.

\begin{figure}
	\centering
	\includegraphics[width=0.9\columnwidth]{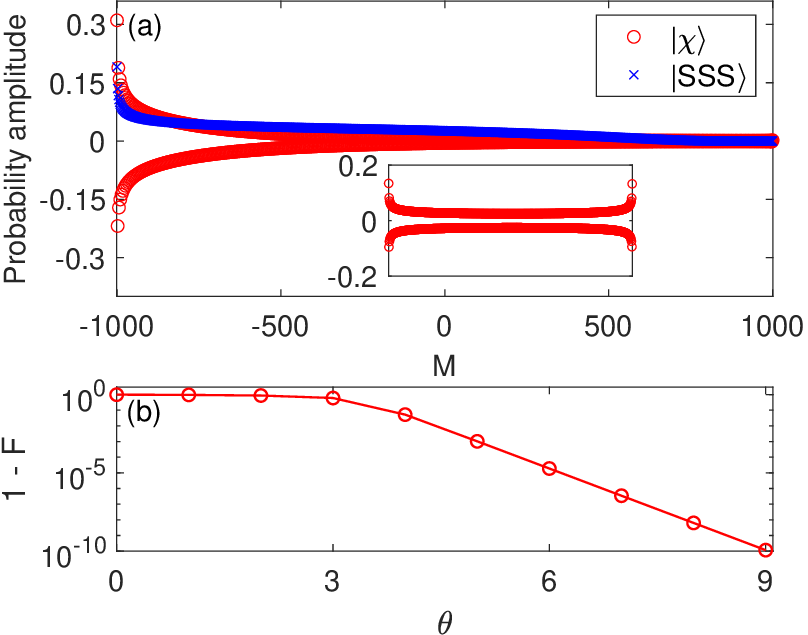}
	\caption{(a) Probability amplitudes of the ground state $\ket{\chi}$ (red circles) and $\ket{SSS}$ (blue crosses) in the basis of $\{\ket{J,M}, M=-J, -J+2, \cdots, J\}$. The odd-$M$ amplitudes are zero (not shown). The parameters are $J=1000$ and $\theta =3$. The inset shows the probability amplitude for $\theta = 9$. (b) Infidelity of the state $\ket{\chi}$ and a Dicke $\ket{\psi_D}$. Obviously the spin squeezed ground state of the OAT Hamiltonian is a Dicke state.}
	\label{fig.prob}
\end{figure}

A typical ground state of the Bogoliubov Hamiltonian $\ket{\chi}$ is presented in Fig.~\ref{fig.prob}(a) for $\theta=3$. As a comparison, the optimal spin squeezed state $|SSS\rangle$ under the TAT Hamiltonian is also shown. Both $|\chi\rangle$ and $|SSS\rangle$ are real function. Besides the odd $M$ coefficients $C_M$ being zero, the even $M$ coefficients of $|SSS\rangle$ are positive while that of $|\chi\rangle$ are alternatively positive and negative. Clearly, these two states are very different, though the spin squeezing parameter $\xi_R^2$ of them is close to the HL. As shown in the inset, the state $\ket{\chi}$ at $\theta = 9$ is similar to a Dicke state $\ket{\psi_D}$ with $J_x\ket{\psi_D} = 0$. We then plot the infidelity of the state $\ket{\chi}$ and $\ket{\psi_D}$ in Fig.~\ref{fig.prob}(b) with $1-F = 1-|\bra{\chi}\psi_D\rangle|^2$. As $\theta$ increases to a large value, the infidelity approaches to zero exponentially, indicating the spin squeezed ground state $\ket{\chi}$ of the OAT Hamiltonian is in fact a Dicke state.

Here we note that the spin squeezing parameter $\xi_R^2$ of the Dicke state was previously considered illy defined because its spin average is zero and other parameters, like Dicke squeezing parameter and Mach-Zehnder phase sensitivity~\cite{Meystre2004PRA, Pezz2005PRA, MA2011,Zhang_2014NJP}, were introduced to characterize the strong entanglement of the state. These parameters exhibit Heisenberg scaling. However, the spin squeezing parameter $\xi_R^2$ is in fact well-defined and converges to the Heisenberg limit $1/J$, even though the spin average is zero~\cite{RMPSmerzi, two_mode}. Such a finite (nonzero) spin squeezing parameter $\xi_R^2$ is due to the exactly same asymptotic behavior of the nominator, $2J(\Delta J_{n_\perp})^2$, and the denominator, $|\langle J_z\rangle |^2$, as $\exp(-2\theta)$ approaches zero (see Eqs.~(\ref{Eq:xi_rs}-\ref{Eq:limval})). In addition to investigating the squeezing and entanglement properties of squeezed states, one may also employ many-body correlators to explore the many-body non-locality of such states~\cite{Plodzie2022PRL}.

\section{Possible experimental platforms}

There are a variety of experimental platforms to potentially realize the HL spin-squeezed ground state $\ket{\chi}$. First, a dipolar spinor BEC is an ideal test bed of the LMG model, where the atoms macroscopically occupy internal hyperfine states that can be treated as collective spin states. Under single mode approximation, the spin-dependent effective Hamiltonian of the condensate with magnetic dipole-dipole interaction in an external magnetic field along $z$-direction is~\cite{Law98,Yi2004PRL,Yi2006PRA,Yi2012PRA}
\begin{equation} \label{Eq:Hameff}
	H_{e}= -D J_z^2+E(J_x^2-J_y^2)+\Omega J_z,
\end{equation}
where ${\mathbf{J}}=\sum_{\alpha\beta}\hat{a}_{\alpha}^{\dagger} \mathbf{F}_{\alpha\beta}\hat{a}_{\beta}$ is the collective condensate spin operator and $J_\eta \; (\eta=x, y, z)$ its $\eta$-component. The constants $D$ and $E$ depends on the density and geometry of the condensate and $\Omega$ on the external magnetic field. By tuning the isotropic parameter $\gamma=(D-E)/(D+E)$ and the field $\Omega = 0$, this Hamiltonian takes the form of OAT model, i.e. $H_{OAT} \approx -D J_z^2$ for $\gamma \approx 1$ and $H_{OAT} \approx 2E J_x^2$ for $\gamma \approx 0$, and TAT model $H_{TAT} \approx E(J_x^2-J_y^2)$ for $\gamma = -1$, or $H_{TAT}\approx 2E(J_x^2-J_{z}^2)$ for $\gamma = 1/2$. We have omitted the constant term proportional to ${\mathbf{J}}^2$.

Second, nonlinear atom interaction in a two-component BEC is described by an OAT Hamiltonian $H_{OAT}=\eta \tilde{J_z}^2$, where $\tilde{J}_{x,y,z}$ is the pseudo-spin operator constituted by two different modes of the condensate and $\eta$ the inter-mode interaction strength. The system can be regarded as a BEC with atoms in two internal hyperfine (or Zeeman) states or a BEC in a double-well potential. The interaction between the pseudo-spin modes for both cases is fine tunable in experiment~\cite{Zoller2001nature,Zoller2002PRA,You2002PRA, Zoller2003PRA,Meystre2004PRA,Pezz2005PRA,Choi2005PRA,You2007PRA}.

Third, both OAT and TAT Hamiltonians may be experimentally realized in a system of spins in a cavity, where the interaction between the spins and the cavity is described by the Hamiltonian $H_{TC}=\omega_m\hat{a}^{\dagger}\hat{a}+\omega_B J_z+g(\hat{a}^{\dagger}J_{-}+\hat{a}J_{+})$, where $\hat{a} (\hat{a}^{\dagger})$ is the annihilation (creation) operator of the cavity field with a mode frequency $\omega_m$, $J_{\alpha} (\alpha=z,\pm$) the collective spin operator, $\omega_B$ the Zeeman splitting of the spins, and $g$ the interaction strength. After Schrieffer-Wolf transformation $e^{R}H e^{-R}$ with $R=(g/\Delta_m)(\hat{a}^{\dagger}J_{-}-\hat{a}J_{+})$, $\Delta_m=|\omega_B-\omega_m|$, the Hamiltonian of the spin part is approximated to $H_{OAT}\approx (2g^2/\Delta_m)J_z^2$, indicating that the effective spin interaction is mediated by the cavity mode~\cite{Agarwal1997PRA,Dimer2007PRA,Morrison2008PRA, Morrison2008PRL, Leroux2010PRL, Bennett2013PRL, Dalla2013PRL,Masson2017PRL,  Zhiqiang2017OPtica, Borregaard2017NJP, Groiseau2021PRA,Groszkowski2022PRL,Li2022PRXQuantum}. When the system subjects to a parametric two-photon driving, the spin interaction term becomes a TAT Hamiltonian $H_{TAT}\approx[2\lambda g^2/(\Delta_m\Delta_c)] (J_x^2-J_y^2)$ with $\lambda$ being the driving amplitude and $\Delta_c$ the detuning between the cavity and driving frequency~\cite{Groszkowski2020PRL}.

Fourth, ultracold atoms loaded in an optical lattice is described by Bose(Fermi)-Hubbard model in which the site-dependent spin operator can be reduced to a collective spin operator in Mott phase. Both OAT and TAT Hamiltonian may be induced by applying an additional weak light to the system~\cite{Sorensen1999PRL,Gerbier2006PRL,MA2011, Plodzien2020PRA,Yanes2022PRL,Dziurawiec2023PRA}.
The effective Hamiltonian depends on the phase $\phi$ of the light, and becomes an OAT Hamiltonian $H_{OAT}\approx\mp\hbar\chi_{\phi}J_{x,z}^2$, for $\phi=\pi, 2n\pi/N$ with $n=\pm1,\pm2,\cdots\pm(N/2-1)$, where $N$ is the atom number and $\chi_{\phi}$ the effective coupling strength. When the system is driven by two lights, the effective Hamiltonian becomes a TAT model $H_{TAT}\approx\hbar\chi_{\phi}J_{z}^2-\hbar\chi_{\pi}J_{x}^2$. This approach is also suitable for preparation of squeezed state in Rydberg atom arrays
 ~\cite{Pedrozo2020Nature,Bornet2023Nature}.

Fifth, for two species of atoms coupled through dipole-dipole interaction in a vapor cell (represented by their collective spin operator $S$ and $J$), periodically driving the system can transform the inter-species dipolar interaction into a nonlinear intra-species interaction, resulting in both effective OAT and TAT Hamiltonians~\cite{Huang2021NPJ,Huang2023PRA}. As for a continuous driving, the OAT Hamiltonian is realized as $H_{OAT}=\chi_{eff}S_z^2$, where $\chi_{eff}=-g^2/(2\Delta_{f})$ with $g$ being the dipolar coupling strength between two species and $\Delta_{f}$ the difference between the magnitudes of two external DC fields applied to the two species of spins respectively. Realization of a TAT Hamiltonian needs an extra AC field applied to the $S$ species, which yields $H_{TAT}=\chi_{eff}(S_x^2-S_y^2)$.

\section{Conclusion}
 We construct a diagonal bilinear spin Bogoliubov Hamiltonian, which includes the one-axis twisting Hamiltonian as a limiting case, by employing the spin lowering and raising operators. We prove analytically and numerically that the ground state of the spin Bogoliubov Hamiltonian (and the one-axis-twisting Hamiltonian) exhibits Heisenberg-limit spin squeezing, $\xi_R^2 \propto J^{-1}$ for an arbitrary spin $J$ in a certain parameter regime, in contrast to previous scaling $\xi_R^2 \propto J^{-2/3}$ under the one-axis-twisting Hamiltonian by quench dynamics. Such a spin-squeezed ground state may be realized experimentally in dipolar spinor BECs, ultracold atoms in an optical lattice, and spins in a cavity or in a vapor cell.\\


\acknowledgements{This work is supported by the National Natural Science Foundation of China (Grant No. 12274331), Innovation Program for Quantum Science and Technology (Grant No. 2021ZD0302100), and the NSAF (Grant No. U1930201). The numerical calculations in this paper have been partially done on the supercomputing system in the Supercomputing Center of Wuhan University.}


\begin{thebibliography}{88}
\bibitem [{app()}]{approx}%
	\BibitemOpen
	\href@noop {} {}\bibinfo {note} {At a large enough $\theta$, $\tanh\theta$
		can firstly be approximated by $(1-2\exp(-2\theta))$. Furthermore, the
		approximation $(\nu^*/\mu)^K \approx 1-2K\exp(-2\theta)$ is valid when
		$\exp(-2\theta)\ll 1/(2K-2)$. Since $\max {K}=J$, the minimum requirement for
		this approximation to be valid for all $K$s is $ \exp(-2\theta)\ll1/(2J-2)$.
		Specifically, for $J=1000$, the valid regime for the the exponential decay
		formula of spin average(Eq.(13)) is $\theta>3.8$, which coincides with the
		results shown in Fig.1(a).}\BibitemShut {Stop}%
\bibitem [{not()}]{note1}%
	\BibitemOpen
	\href@noop {} {}\bibinfo {note} {We note that the phase sensitivity is in
		principle determined by the quantum Cramer-Rao bound
		$\Delta\phi=\xi_R/\sqrt{J}\geq1/\sqrt{F_Q}$, where $F_Q$ denotes the quantum
		Fisher information. From this inequality, one finds that the spin squeezing
		is limited by $F_Q$. It is evident that the GIS $\ket{\chi}$ saturates the
		quantum Cramer-Rao bound, resulting in
		$\xi_R^2=2J/F_Q$~\cite{Ma2009PRA}.}\BibitemShut {Stop}%
\bibitem [{two()}]{two_mode}%
	\BibitemOpen
	\href@noop {} {}\bibinfo {note} {The same effective Hamiltonian $H_A$ is
		derived approximately in a Bose-Einstein condensate trapped in a double-well
		potential and the Heisenberg-limit spin squeezing may be achieved under the
		two-mode, large $J$, and mean-field approximations. However, our results on
		the spin Bogoliubov Hamiltonian $H_B$ are free of these approximations thus
		are applicable to many other systems.}\BibitemShut {Stop}%
\end{thebibliography}
%

\end{document}